\documentclass[a4paper]{article}

\usepackage{amsmath}
\usepackage{icrc2013}
\usepackage{indentfirst}

\title{Why should we keep measuring zenital dependence of muon flux? Results obtained at Campinas (SP) BR}

\authors{
B.Daniel,
L. M. Santos,
M. Nunes,
T. V. Vieira,
E. Kemp
}

\afiliations{
\scriptsize{
UNICAMP - State University of Campinas
}
}

\email{bdaniel@ifi.unicamp.br}

\abstract{The zenital dependence of muon flux which reaches the earth's surface is well known as proportional to $cos^{n}(\theta)$.
Generally, for practical purposes and simplicity in calculations, n is taken as 2. However, compilations of measurements 
show dependence on the geographical location of the experiments as well as the muons energy range. Since analytical
solutions appear to be increasingly less necessary because of the higher accessibility to low cost computational power,
accurate and precise determination of the value of the exponent n, under different conditions, can be useful in the
necessary calculations to estimate signals and backgrounds, either for terrestrial and underground experiments. In this
work we discuss a method for measuring n using a simple muon telescope and the results obtained for measurements
taken at Campinas (SP), Brazil ($22^o$ 54' W, $-41^o$ 03', 854 m asl). After validation of the method, we intend to extend the
measurements for different geographic locations due to the simplicity of the method, and thus collect more values ​of n
that currently exist in compilations of general data on cosmic rays.
}


\begin{document}
\maketitle


\section{Introduction}
 
Muons represent 10\% of the energy of an extensive air shower and because of their heavy mass and relativistic behavior, are the most abundant
particles at ground level produced by the cosmic radiation. Therefore, they allow the indirect detection of cosmic rays in the surface, trough the
reconstruction of the air shower, using data took by different detectors working in time coincidence to infer the direction of the primary particle. This
technique was successfully used in many experiments, like the Pierre Auger Observatory \cite{bib:auger} for ultra-high energies. Muons also represent a source
of noise in other particle physics experiments, like underground neutrino detectors. For these reasons it is very important to study their flux at the surface.

The muon flux has a peculiarity: it describes an anisotropic angular distribution that goes with a $cos^{n} \theta$, $\theta$ being the zenith angle.
This feature was widely discussed, especially the parameter $n$ which was established \cite{bib:allkofer} as 2. But there is no reason for this assumption
to be generalized. This value for the parameter may depend on the region of the planet, altitude and period of the measurements, and also be related to the
energy range considered.

This work proposes a simple experimental method to measure $n$, as a first step to study the muon flux in details and check the dependences
aforementioned. It makes use of a muon telescope made of three paddle shaped scintillator detectors, vertically aligned, operating in time coincidence.
Counting rate measurements are made for different distances between the outermost paddles. Each distance defines an aperture for the telescope, which is
related to the geometry of the telescope and the zenith dependence of the flux of incoming particles.

In next session, a theoretical approach to the telescope aperture $\Gamma$ and the muon flux are set forth. Section 3 shows the method developed to obtain
$n$. Finally, section 4 describes the experimental setup used and the results obtained.

\section{Aperture of a particle telescope}

 A vertical arrangement of detectors, at a fixed distance between them, defines an acceptable range of zenital angles that grants the collection of particles striking the detectors.
This is given by the \textsl{aperture}\footnote{Other acceptable terms are \textsl{gathering power} and \textsl{angular acceptance}} of the telescope, $\Gamma$.
To define it we must write an expression for the counting rate of the telescope \cite{bib:sullivan}:
\begin{eqnarray}
 \label{eq01}
 C(\vec{x},t_{0}) = \int^{t_{0}+T}_{t_{0}} dt \int_{S} d\vec{\sigma} \cdot \hat{r} \int_{\Omega} d\omega \int^{\infty}_{0} dE \times \nonumber \\
\times \sum_{\alpha}\epsilon_{\alpha}(E,\vec{\sigma},\omega,t)\vec{J}_{\alpha}(E,\omega,\vec{x},t)
\end{eqnarray}
where 
\begin{itemize}
 \item $C$ is the counting rate in $s^{-1}$ during a time interval $t$ between $t_{0}$ and $t_{0} + T$,
 \item $J_{\alpha}$ and $\epsilon_{\alpha}$ are, respectively, the spectral intensity and the detection efficiency
 of the $\alpha^{th}$ particle in $s^{-1} cm^{-2} sr^{-1} E^{-1}$,
 \item $d\vec{\sigma}$ is the infinitesimal element of area of the last sensor with total area $S$,
 \item $d\omega$ is the infinitesimal element of solid angle inside the domain $\Omega(\theta,\phi)$,
 \item $\vec{x}$ is the spatial coordinate of the telescope,
 \item $\hat{r}$ is the unit vector in the direction of $\omega$, and
 \item $\hat{r} \cdot d\vec{\sigma}$ is the effective element of area looking into $\omega$. 
\end{itemize}

Although equation (\ref{eq01}) is quite general, there are still some assumptions that must be
taken into account, like
\begin{enumerate}
 \item $d\vec{\sigma}, \omega$ and $\vec{x}$ are all time independent, because the telescope is at rest;
 \item There are not any decays or transformation of particles through the passage by the detector,
 unless it is specified in $\epsilon_{\alpha}$;
 \item There is not any kind of process that could deviate the particle linear trajectory;
 \item $J_{\alpha}$ is independent of $\sigma$ and $\epsilon_{\alpha}$ is independent of $\vec{x}$.
\end{enumerate}

To further simplifications lets consider an ideal telescope with its efficiency independent of
$\omega, \sigma$ and $t$ leading to
\begin{eqnarray}
  \label{eq02}
  \epsilon_{\alpha} = 0, & \alpha \not= 1 \nonumber \\
 \epsilon_{1} = 1, & E_{i}\le E \le E_{f} \\
 \epsilon_{1} = 0, & otherwise. \nonumber
\end{eqnarray}
Henceforth, the subscript can be dropped. Assuming that the spectral intensity of the particle has no dependance
in $\vec{x}$ and $t$, it can be rewriten as
\begin{eqnarray}
\label{eq03}
\vec{J}(E,\omega) = J_{0}(E) K(\omega)
\end{eqnarray}
which leads to a simplification of equation (\ref{eq01})
\begin{eqnarray}
 \label{eq04}
 C = \left[\int_{\Omega} d\omega \int_{S}d\vec{\sigma}\cdot\hat{r}K(\omega)\right] \int_{E_i}^{E_{f}} dE J_{0}(E).
\end{eqnarray}

The aperture $\Gamma$ is then defined as the term between the square brackets in equation (\ref{eq04})
 \begin{eqnarray}
 \label{eq05}
 \Gamma \equiv \int_{\Omega} d\omega \int_{S}d\vec{\sigma}\cdot\hat{r}K(\omega) \nonumber \\
	= \int_{\Omega} d\omega K(\omega) D(\omega) 
\end{eqnarray}
where $D(\omega)$ defines the \textsl{Directional Response Function}
\begin{eqnarray}
\label{eq06}
D (\omega) \equiv \int_{S} d\vec{\sigma} \cdot \hat{r}.
\end{eqnarray}

For rectangular detectors which defines a axis-simetrical telescope, the counting rate
for individual particles is given by
\begin{eqnarray}
 \label{eq07}
 \frac{dN}{dt} = \int_{\Omega} d\omega I(\omega) D(\omega)
\end{eqnarray}

\begin{figure}[h]
  \centering
  \includegraphics[width=9cm]{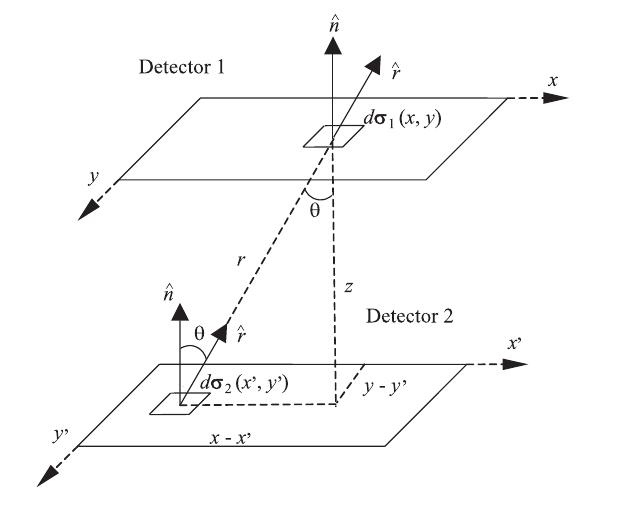}
  \caption{Schematic arrangement of detectors for the calculation of aperture. Extracted from \cite{bib:rbef}.}
  \label{fig:aperture}
\end{figure}

Assuming that the radiated particles have an angular distribution $I(\omega) = I_{0} K(\omega) = I_{0}cos^{n} \theta$, $cos \theta = Z/r$ (figure
\ref{fig:aperture}), then equation (\ref{eq07}) can be
rewrited as 
\begin{eqnarray}
 \label{eq08}
 \frac{dN}{dt} = I_{0}\int_{\Omega}cos^{n} \theta d\omega \int_{S} d\vec{\sigma}\cdot\hat{r}  = I_{0} \Gamma_{n} 
\end{eqnarray}
where $\Gamma_{n}$ is the already defined aperture. Therefore, the counting rate is proportional to the telescope geometry with $I_{0}$ defining the vertical
flux of particles passing through it.

The integration of $D(\omega)$ (equation(\ref{eq06})) for an element of area $d\sigma_2 = dx\prime dy\prime$
(figure \ref{fig:aperture}), returns the directional response in function of the detector's geometrical parameters X, Y and Z:
\begin{eqnarray}
\label{eq09}
 D (\omega) = \int_{S} d\vec{\sigma} \cdot \hat{r} = \int cos \theta dx\prime dy\prime \nonumber \\
   = cos \theta(X - |Ztan \theta cos \phi|)(Y - |Ztan \theta sin \phi|)
\end{eqnarray}
Applying equation (\ref{eq09}) in (\ref{eq07}), an expression for the counting rate in the $\omega$ direction is defined:
\begin{eqnarray}
 \label{eq10}
 \frac{d^{2}N}{d\omega dt} = I_{0} = cos^{n+1} \theta(X-|\xi|)(Y-|\eta|)
\end{eqnarray}
where
\begin{eqnarray}
  \label{eq11}
 \xi \equiv Z tan \theta cos \phi,&(-X \le \xi \le X) \nonumber \\
 \eta \equiv Z tan \theta sin \phi,&(-Y \le \eta \le Y) 
\end{eqnarray}
Therefore, using equation (\ref{eq07}), a final expression for the aperture $\Gamma_{n}$ can be obtained
\begin{eqnarray}
 \label{eq12}
 \Gamma_{n} = \frac{1}{Z^{2}}\int_{-X}^{X}\int_{-Y}^{Y}cos^{n+4}(X-|\xi|)(Y-|\eta|)d\xi d\eta \nonumber \\
 = \frac{4}{Z^{2}}\int_{0}^{X}\int_{0}^{Y}cos^{n+4}(X-\xi)(Y-\eta)d\xi d\eta \nonumber \\
 = 4Z^{n+2} \int_{0}^{X}\int_{0}^{Y}\frac{(X-\xi)(Y-\eta)}{[Z^{2}+\xi^{2}+\eta^{2}]^{(n+4)/2}}d\xi d\eta.
 \end{eqnarray}

\section{Experimental methodology}

\subsection{Experimental setup}

The muon telescope used consists of three detectors of plastic scintillator (0,4 x 0,4 x 0,05 $m^3$) with light guides coupled to photomultipliers (PMTs). Each
detector is connected to a high voltage power supply adjusted to its optimum value (the one that ensures maximum efficiency of the PMT). Particles crossing
the scintillators generate light pulses converted to analogical electric signals, processed with NIM and CAMAC standard modules, for discrimination, time coincidence and counting. 

The rate of single counting, double (considering the outermost detectors) and triple coincidences were recorded every ten minutes for a period of 4
hours, giving a set of measurements for each height of the telescope, ranging from $0.3$ to $2.1$ meters. The efficiency $\epsilon$ of the telescope was evaluated taking the ratio $\epsilon = N_{T}/N_{D}$ between triple and double coincidences for each distance. All frequency data were normalized considering the efficiency of each detector. We have to mention that during data taken the atmospheric pressure was monitored, but no significant modulations of the telescope counting rate was observed in correlation with amplitude of pressure variations. 

The same procedure described above was done putting a layer of 5 cm of lead above the last detector (closer to the ground), to select particles with energy $E > 100$ MeV.

\subsection{Estimation of the parameter $n$}

Given the dimensions of a telescope (X, Y and Z), the aperture $\Gamma_n$ can be calculated with equation (\ref{eq12}) for any value of $n$. If $\Gamma_n$ is
obtained for different values of Z (the distance between the outermost detectors) and the counting rate of particles is also measured, it is possible to obtain
the muon flux trough relation (\ref{eq08}), with a linear fit from counting rate vs. aperture data.

\begin{figure}
\center
\scalebox{0.4}{\includegraphics{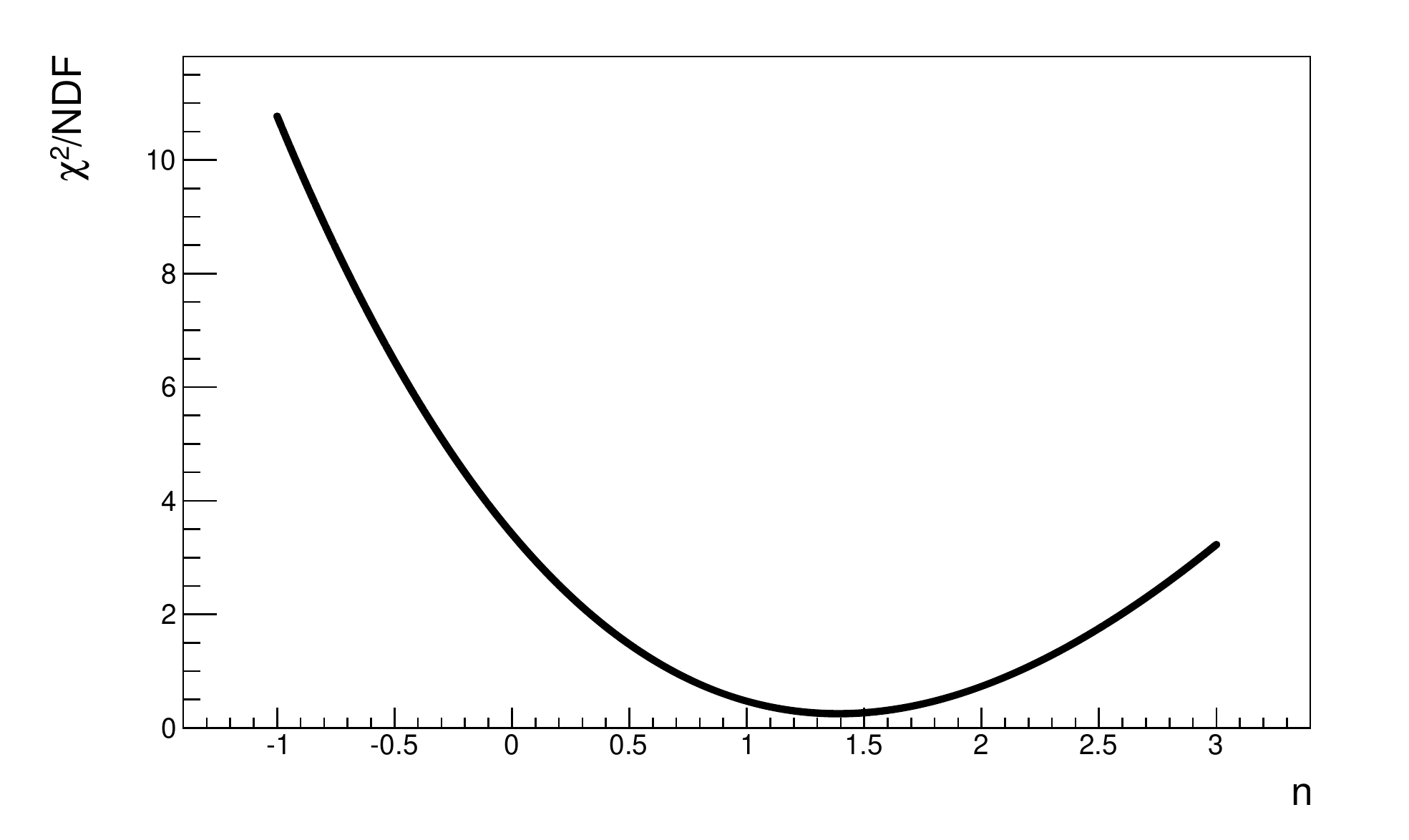}}
\caption{$\chi^2/NDF$ obtained for different values of $n$.}
\label{chi2}
\end{figure}

The parameter $n$ can be estimated repeating this previous calculation for different values of $n$ and evaluating the reduced $\chi^2$, normalized by the number of degrees of fredom ($\chi^2$/NDF), for each value of $n$. Figure \ref{chi2} shows results for 5000 random values of $n$, selected within a suitble range, and the calculated $\chi^2/NDF$ for the telescope described in the previous section. The best value for $n$ is chosen as the one which minimizes the $\chi^2/NDF$ and its deviation was defined by the $68\%$ of confidence level region.

\section{Results}

\begin{figure}[!h]
  \centering
  \includegraphics[width=8cm]{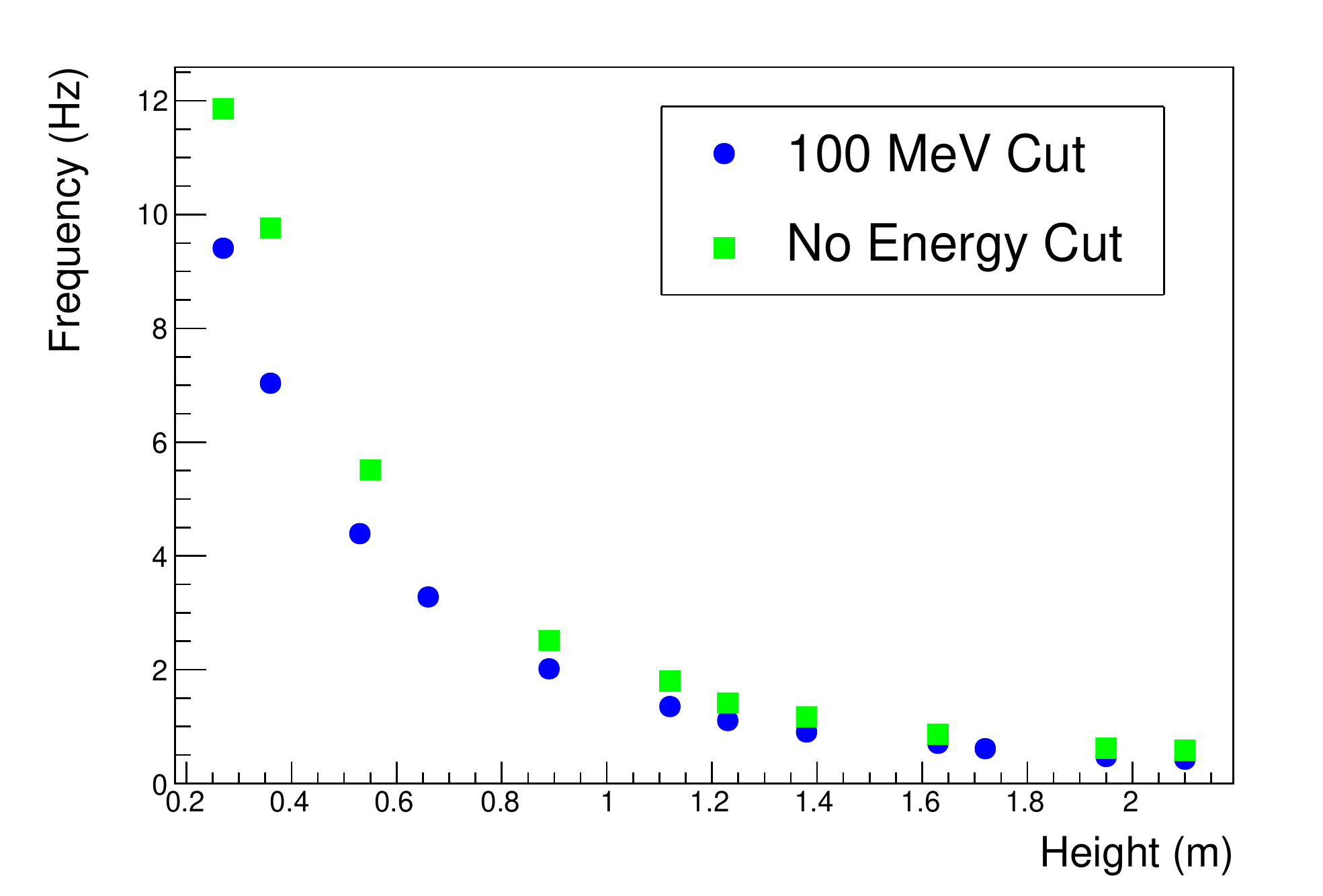}
  \caption{Data taken with the muon telescope described in section 4, with the lead layer (blue dots) and without an energy cut (green squares).}
   \label{h_c}
\end{figure}

Figure \ref{h_c} shows the results obtained for both telescope assemblies, with no energy selection and with the lead absorber. The non-linear dependence of the counting frequency on telescope sensors separation is in agreement with the expected values from equations \ref{eq08} and \ref{eq12}. It can also be seen that there is a global down-shift in data from the case we used the lead absorber, thus consistent with the energy cut applied.

\begin{figure}[!h]
  \centering
  \includegraphics[width=8cm]{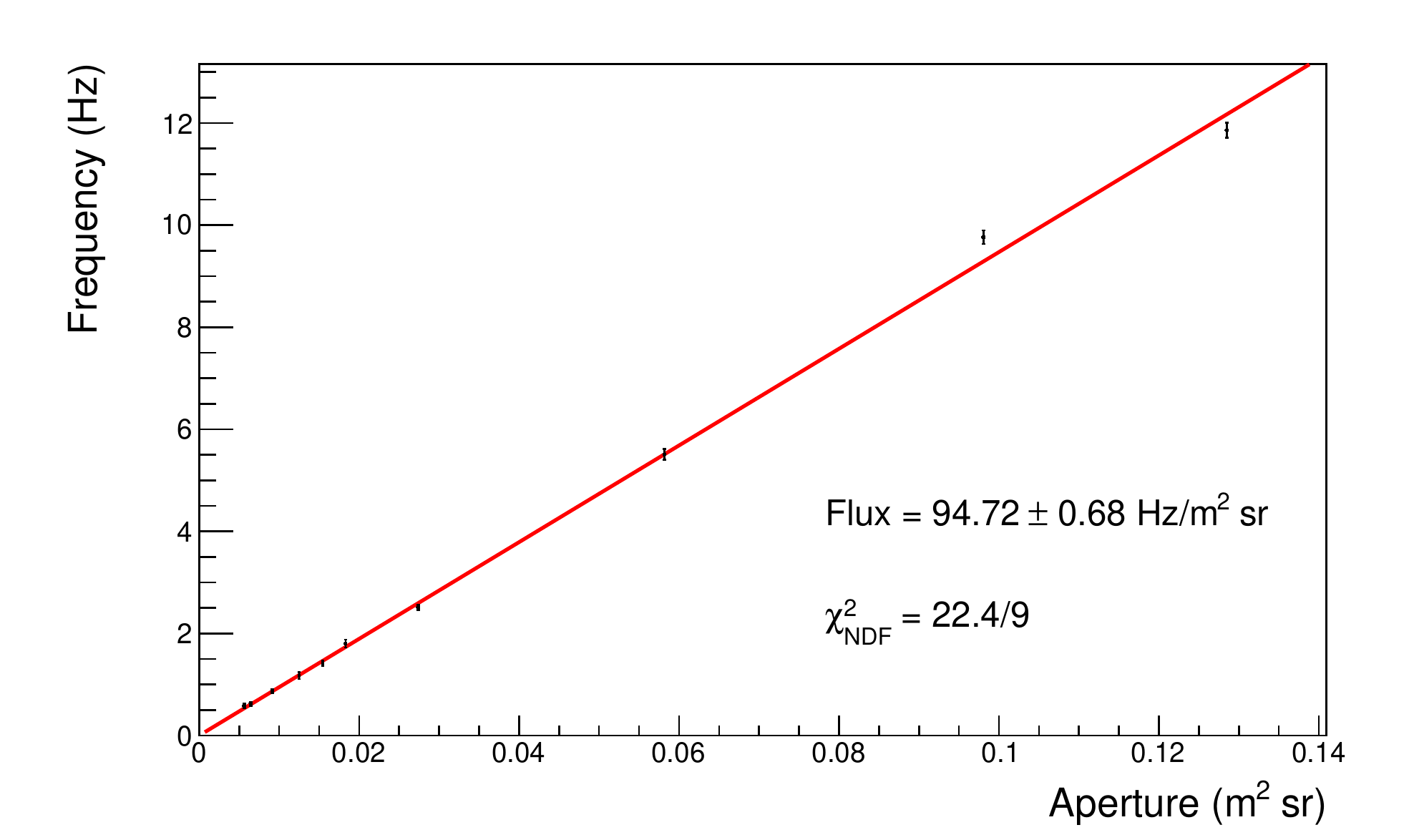}
  \caption{The best fit for the frequency dependence with the telescope aperture without a lead absorber.}
  \label{ap_s}  
\end{figure}

\begin{figure}[!h]
  \centering
  \includegraphics[width=8cm]{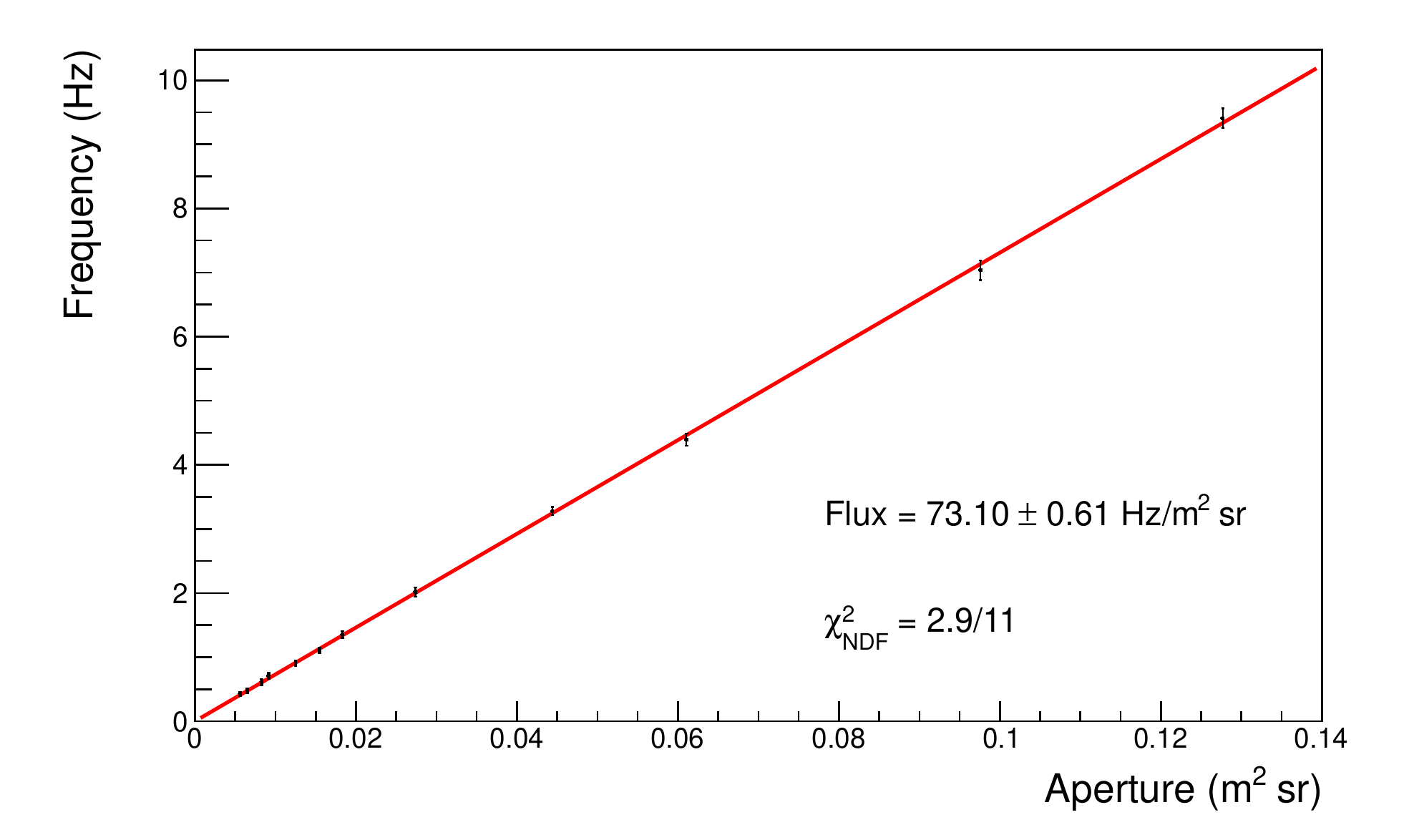}
  \caption{The best fit for the frequency dependence with the telescope aperture with a lead absorber.}
   \label{ap_c}
\end{figure}

Following the method described in section 3, the value of $n$ which minimizes the $\chi^2/NDF$ of the linear fit from frequency vs. aperture was calculated for data acquired with both assemblies. These fits are shown in figures \ref{ap_s} and \ref{ap_c} for the optimum value of $n$. The results for $n$ and the flux $\Phi$ are
summarised in table \ref{table_conc}. As mentioned before, deviations in $n$ were determined by the $68\%$ of confidence level region whereas the flux deviations were obtained straight from the error for the angular coefficient of the linear fit.

\begin{table}[!h]
\centering
\caption{Values of $n$ and Flux($\Phi$) for each assembly.}
\label{table_conc}
\begin{tabular}{cccc}
  \hline
  Energy Cut & $n$ & $\Phi (Hz/m^{2} sr)$ & $\chi^2/NDF$ \\
  \hline
  No cut & $(1.46 ^{+0.19}_{-0.18})$ & $(94.72 \pm 0.68)$ & $22.4/9$ \\
  
  100 MeV & $(1.51 ^{+0.16}_{-0.15})$ & $(73.10 \pm 0.61)$ & $2.9/11$ \\
  
 \end{tabular}
\end{table}

These results are consistent  with other measurements \cite{bib:turco} and also shows self-consistence. Some discrepancy between the vertical flux quoted in this work and a previous one measured at the same location \cite{bib:rbef} have to be cross-checked with expected variations due the solar cycle.

\section{Summary}

Generally, the muon signals and backgrounds in particle physics experiments are estimated assuming the index $n$ of zenith angle dependence as $n = 2$, but $n$ may depend on the location of the measurements and other conditions. A rather simple and fast method to determine local values for the $cos^n$ dependence of the muon flux in the surface was described. It can be done with a telescope with variable distance between the outermost detectors and not requires any further inclined detector arrangement.  With this simple method, it is possible to simultaneously measure $n$ and the vertical flux, avoiding assumptions about generic and/or typical values.

Our measurements of the muon flux in Campinas, Brazil, with plastic scintillator detectors, result in the following values: $\Phi = (94.72 \pm 0.68) Hz/m^{2} sr$, and $n = (1.46^{+0.49}_{-0.46})$, in case of no energy cuts. The energy selection $E > 100 MeV$ by adding a 5 cm layer of lead to the lowest level sensor of the telescope, resulted in $\Phi = (73.10 \pm 0.61) Hz/m^{2} sr$ and  $n = (1.51 ^{+0.65}_{-0.60})$ . This experimental results are compatible with other measurements \cite{bib:turco}, and are different from the expected value for $n$, which is usually taken as $n = 2$ as discussed in the text.

These measurements can be easely performed in different locations, providing a broader range of physical measurements beyond the muon flux intrinsic features such as zenith angle dependence or vertical intensity. Possible observations in variations of these parameters can be related to height, latitude, pressure and energy cut-off effects induced by phenomena like the geomagnetic anomaly, injection of coronal mass, and changes in the solar activity cycle. In a future work, a compilation of other results will be used for this purpose.

\footnotesize{{\bf Acknowledgment:} { The colaborators thank CAPES, FAPESP and APGF for several financial supports.}}

\end{document}